\documentclass[aps,prb,twocolumn,showpacs,amsmath,amssymb]{revtex4}
\usepackage{epsfig}
\usepackage{dcolumn}
\usepackage{bm}
\usepackage{amsmath}
\usepackage{amsfonts}
\usepackage{amssymb}
\usepackage{graphicx}%
\newcommand{\beq}{\begin{eqnarray}}
\newcommand{\eeq}{\end{eqnarray}}

\begin{document}
\title{ac-dc voltage profile and four point impedance of a quantum driven system}
\author{Federico Foieri and  Liliana Arrachea.}
\affiliation{Departamento de F\'{\i}sica ``J. J. Giambiagi" FCEyN, Universidad de Buenos 
Aires and Instituto de F{\'i}sica de Buenos Aires, 
Ciudad Universitaria Pab.I, (1428) Buenos Aires, Argentina.}
\date\today 
\begin{abstract}

We investigate the behavior of the time-dependent voltage drop in a periodically driven 
quantum conductor sensed by weakly coupled dynamical voltages probes.
 We introduce the concepts of ac-dc local
voltage and four point impedance in an electronic system driven by ac fields.
We  discuss the properties of the different components of these
quantities in a simple model of a quantum pump, where  two ac voltages oscillating with 
a phase lag are applied at the walls of a quantum dot. 
\end{abstract}
\maketitle
\section{Introduction}\label{int}

In stationary transport through mesoscopic systems, the four-point  terminal resistance is regarded as
the proper concept to characterize the resistive behavior of the sample, free from the effects of the
contact resistances. This concept has been introduced in Refs. \onlinecite{L70,engq}, and further
elaborated in Refs. \onlinecite{BU8688,been}. Its behavior in different systems under dc-driving has been
analyzed in several theoretical works. \cite{gram} Recent experiments on semiconducting devices 
\cite{pic} and carbon nanotubes \cite{gao,makarovski} constitute evidences that 
this quantity can be positive as well as negative at low temperatures, in agreement with 
  theoretical predictions on the basis of coherent electronic transport.
\cite{L70,engq,BU8688,been,gram}

Time dependent quantum transport in ac driven small-size systems is receiving nowadays considerable theoretical
and experimental attention. A variety of devices like quantum dots, 
electronic systems in the quantum Hall regime, quantum capacitors and graphene nanoribbons have
 been recently
investigated experimentally. 
\cite{SMCG99,Chepe,pumpex} 
Among other interesting effects, 
 the mechanisms for the induction of dc electronic and spin currents 
\cite{tdep1_spin_charge,mobum,brou0,brou,mobup,lilimos,lilip,lilisim,otros}, 
the behavior of the dc and t-resolved noise \cite{tdep2_noise1,tdep2_noise}, the energy transport 
and the heat generation \cite{tdep3_heat}   have been  analyzed. 

While the experimental setups in some of these devices
 involve four-terminal measurements \cite{SMCG99,Chepe}, the theoretical discussion 
on how to extend the definition of the  four-point resistance in the context of
 time-dependent transport, has been 
considered only recently. In Ref. \onlinecite{us}  
we have introduced the concept of  a non-invasive  dc voltage probe in a simple model of a 
quantum pump.
We have extended the ideas of Refs. \onlinecite{L70,engq,BU8688,been,gram} by representing the probe
as a particle reservoir which is weakly coupled to the driven system at the point where the 
voltage
is to be sensed. Then, the  {\em local dc voltage} is defined as the  value of the dc bias
that has to be applied at the probe in order to satisfy the condition of
a vanishing dc particle current between it and the driven system. A similar route has been recently
 followed to define the local temperature
from the constraint of a vanishing dc heat current between the driven system and the probe, as
a condition of local thermal equilibrium. \cite{temp}
The dc four point resistance is defined as the ratio between  
the  dc voltage drop measured between   two independent weakly coupled voltages probes   
and the dc  
pumped current circulating through the device. \cite{us} In this {\em ``gedanken''} setup,
the two probes correspond to sensing the voltage difference between two
points of the circuit by means of a dc voltimeter.

In the presence of ac fields, it is however interesting
 to characterize not only the dc  but also the ac component of
the voltage drop. The aim of this work is precisely to discuss the way to generalize 
the properties of a voltage probe in order 
to sense both the dc and the ac features  in the voltage profile.
Following this route, we are lead to  the  concept of   
{\em four terminal impedance} for a quantum
driven system, as a  concomitant extension of the concept of four terminal resistance. 
 For sake of simplicity, we
 mainly focus on  the  weak driving regime. This corresponds to the so called adiabatic regime,
where the period of the ac voltages is much larger than the typical time that an electron
 spends inside the structure (the dwell time), while the amplitudes of the potentials are much
smaller than the energy scale characterizing the dynamics of the electrons within the structure. 
We also analyze these ideas in a 
simple model of a quantum pump device. 

In Refs. \onlinecite{brou,mobup} it has been 
pointed out that different 
contributions to the dc currents  can be identified in setups under the action of
both localized time-dependent potentials acting on the central structure
 and ac voltages at the reservoirs. The two
most relevant contributions are (i) the one
 due to pure pumping processes, and (ii) the one due to the existence of a bias applied
at the reservoirs. 
The latter part, in turn, may contain a component due to a dc  bias
and a component due to the rectification of the ac potentials. 
Besides these, there is an additional component  in the dc current, which is due to the interference
between the pumping and rectification processes. \cite{mobup,lilimos}
In this work, we show that these different mechanisms affect  
the determination of the local voltage. One remarkable consequence of this fact is 
that the solely introduction of an ac 
voltage at the probe in order to detect time-dependent features at the structure, 
originates additional scattering processes that modify the dc voltage profile, even 
when the probe is 
weakly coupled to the sample.

The paper is organized as follows: In Sec. \ref{s1} the model for the driven 
structure and for  the {\em ac-dc voltage probe} is introduced. We present the 
theoretical treatment based
in non-equilibrium Green functions, used to evaluate the relevant physical 
quantities like the time-dependent currents along the device. 
In Sec. \ref{s2} we present results for the parameters characterizing the 
{\em ac voltage profile}
as well as the four terminal impedance in a model for a quantum pump. 
Finally Sec. \ref{con} is devoted to the summary and conclusions.

\section{Theoretical treatment}\label{s1}
\subsection{Model}\label{model}

\begin{figure}
\includegraphics[width=0.9\columnwidth,clip]{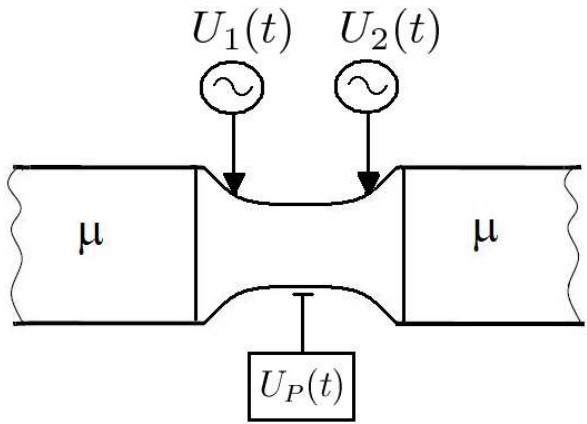}
\caption{\label{fig1} (Color on-line)
Sketch of the setup to describe the ac-dc voltage profile.
 The central system is driven out-of equilibrium by external
ac voltages $U_1(t)$ and $U_2(t)$, while it is in contact to left ($L$) and right ($R$) reservoirs
at the same chemical potential $\mu$ and temperatures $T$. An additional reservoir (the probe)
is weakly coupled to device under investigation. It has a bias with a dc component $\mu_P-\mu$
and an ac voltage $V_P \cos(\Omega_0 t + \varphi_P)$ that locally and instantaneously sets
the equilibrium between the probe and the central system from the condition of
a vanishing instantaneous current between both systems.}
\end{figure}

We consider the setup shown in the sketch of Fig.\ref{fig1}, where a structure
of a finite size  driven by ac potentials 
is in contact to two reservoirs at the same temperature and chemical
potential. For simplicity, we adopt units where $e=\hbar=1$. We describe the system by  the following Hamiltonian:
\begin{eqnarray}\label{hsys}
H_{sys}(t) & = & H_{cen}(t) + \sum_{\alpha=L,R}( H_{\alpha} + H_{c,\alpha}),
\end{eqnarray}
where $L,R$ labels, respectively, left and right reservoirs.
We assume a lattice model with $N$ sites for the central driven system:
\begin{eqnarray}
H_{cen}(t)&= & H_0 + H_{ac}(t), \nonumber\\
H_0 & = &  -w \sum_{ \langle l l'\rangle} [
c^{\dagger}_{l} c_{l'} + H.c.]
+ \sum_{l}^{N} \varepsilon_l c^{\dagger}_{l} c_{l'},
 \nonumber \\
H_{ac}(t)& = &
\sum_{ l,l'} [U_{l,l'}(t) c^{\dagger}_{l} c_{l'}+ H. c.],
\end{eqnarray}
  where $\langle l l' \rangle $ denotes a pair of nearest-neighbor sites,
and $w$ is a hopping parameter, while
$U_{l,l'}(t)=  U^0_{l,l'} \cos(\Omega_0 t + \varphi_{l,l'})$.  
Fig. \ref{fig1} corresponds to an example where this ac potential has two local
components. Assuming that the points of the structure at which the potentials 
act are labeled, respectively, by $l_1$ and $l_2$, the driving potential in this 
example
reads: $U_{l,l'}(t)=  \delta_{l,l'} 
[ \delta_{l,l_1} U^0_1 \cos(\Omega_0 t + \varphi_1) + 
\delta_{l,l_2} U^0_2 \cos(\Omega_0 t + \varphi_2) ]$.

The Hamiltonians for the $L$ and $R$ reservoirs correspond to free electrons:
\begin{eqnarray}
H_{\alpha} & = & \sum_{k_{\alpha}} \varepsilon_{k_{\alpha}} 
c^{\dagger}_{k_{\alpha}} c_{k_{\alpha}}, 
\end{eqnarray}
having chemical potential $\mu$ and equal temperature.
The contacts between the driven system and the reservoirs are described by tunneling
Hamiltonians of the form:
\begin{eqnarray}
H_{c, \alpha} & = & w_{\alpha} \sum_{k_{\alpha}} 
(  c^{\dagger}_{k_{\alpha}} c_{l_{\alpha}} + H.c.), 
\end{eqnarray}
where $l_{\alpha}$ labels the sites of the central lattice that are in contact with
the reservoirs.

We now introduce the model for the ac-dc voltage probe.
It consists in an additional reservoir of non-interacting electrons with a time-dependent
bias voltage, which is weakly coupled to the central device at the point where the potential
is to be sensed. The corresponding Hamiltonian for this system reads: \cite{mobup,lilip,jau}
\begin{eqnarray}
H_{P} & = & \sum_{p}[ \varepsilon_{p} - U_P(t) ]   c^{\dagger}_{p} c_{p} , 
\end{eqnarray}
where $\varepsilon_{p}$ is the dispersion relation corresponding to the free electrons 
while the bias $U_P(t)$ is assumed to depend harmonically in time:
\begin{equation}
U_P(t)= \sum_{k=-\infty}^{+\infty} e^{-i k \Omega_0 t} U_P^{(k)},
\end{equation}
having a dc component $U_P^{(0)}=\mu_P-\mu$ and an 
ac component $U_P^{(ac)}=\sum_{k \neq 0} e^{-i k \Omega_0 t} U_P^{(k)}$.
The probe couples to the central device  at the site $l_P$ through 
a tunneling term of the form:
\begin{eqnarray}
H_{c, P} & = & w_{P} \sum_{p} 
(  c^{\dagger}_{p} c_{l_P} + H.c.).
\end{eqnarray}
We assume that the probe is non-invasive, which implies that the tunneling parameter
$w_P$ is so small that it does not affect the coherent nature of 
the transport processes
along the driven central system. The key feature of the probe is that the  potential $U_P(t)$
is adjusted in order to  satisfy at every time the condition of a vanishing charge current
$J_P(t)=0$ through its contact to the central device (see Fig. \ref{fig1}).
In this way, the  potential $U_P(t)$ is the one satisfying
at every time local equilibrium regarding charge flow between the central system
and the probe. For this reason, it is interpreted as the time-dependent local potential
of the system sensed by the probe. This definition is precisely an extension of the
one originally proposed by Engquist and Anderson \cite{engq} to the case of a system
driven by time-dependent fields. It also generalizes the definition of the dc voltage
probe that we have introduced in Ref. \onlinecite{us}, where we have followed a
procedure equivalent to the present one but with $U_P^{(ac)}=0$. 
As we shall see, to include an ac component
in the probe voltage introduces significant corrections to the sensed dc voltage.

\subsection{Sensing an ac-dc local voltage  with a probe}   \label{volt}
The model for the probe we have introduced in the previous subsection
is completely general. For sake of simplicity, in what follows we focus on weak driving. 
Therefore, we assume that the driving potentials $U(t)$ depend at least on two parameters,
in order to produce adiabatic dc currents at low driving frequencies $\Omega_0$.
\cite{brou,mobup,lilip} We also assume that the corresponding 
driving amplitudes 
are small enough to generate time-dependent currents composed of a single 
harmonic besides the dc component. In particular, 
 we assume that the time-dependent current flowing into the reservoir $\alpha$ 
has the form:
\begin{eqnarray} \label{tdepcur}
J_{\alpha}(t)&=& \sum_{k=-1}^1 J^{(k)}_{\alpha} e^{-i k \Omega_0 t},
\end{eqnarray} 
which motivates assuming the following functional form for the 
ac voltage at the probe:
\begin{eqnarray}
U_P(t)&=& \mu_P- \mu + V_P \cos(\Omega_0 t + \varphi_P).
\end{eqnarray} 
This means that  the local voltage sensed by the probe becomes characterized 
by the dc bias $\mu_P-\mu$, as well as the amplitude $V_P$ 
 and the phase
 $\varphi_P$ of the ac component. These three parameters are adjusted to satisfy
the following set of three equations:
\begin{equation}\label{cond}
J^{(k)}_P  = 0 , \;\;\;\;k=-1,0,1,
\end{equation} 
with $J^{(1)}_P = [J^{(-1)}_P]^*$. 

The evaluation of the different harmonics of the ac current can be done
by resorting to non-equilibrium Green function formalism. Following Ref.
\onlinecite{lilip}, we express the time-dependent current flowing through the 
contact between the central system and the probe in terms of Green functions:
\begin{eqnarray}\label{jpt}
J_{P}(t)&=&  \int_{-\infty}^{+\infty} dt_1  \{ G^R_{l_P,l_P}(t,t_1) \Sigma^<_P(t_1,t)
\nonumber \\
& & + G^<_{l_P,l_P}(t,t_1) \Sigma^A_P(t_1,t) \},
\end{eqnarray} 
with:
\begin{eqnarray} \label{selfp}
\Sigma^A_P(t,t^{\prime}) & = &  i \Theta(t^{\prime}-t) \phi(t,t^{\prime})
 \int \frac{d \omega}{2 \pi}
e^{-i \omega (t- t^{\prime})}\Gamma_P(\omega),
\nonumber \\
\Sigma^<_P(t,t^{\prime}) & = &  i \phi(t,t^{\prime}) \int \frac{d \omega}{2 \pi}
 e^{-i \omega (t- t^{\prime})}
f(\omega)\Gamma_P(\omega),
\end{eqnarray}
with
\begin{eqnarray}\label{phi}
\phi(t,t^{\prime}) &= & e^{-i \int^t_{t^{\prime}} dt_1 U_P(t_1)},
\end{eqnarray}
being $\Gamma_P(\omega) = 2 \pi |w_P|^{2} \sum_p \delta(\omega- \varepsilon_p)$, the
spectral density associated to the self-energy due to the escape of the electrons
from the central device to the probe.
We consider a wide-band model for this system, which implies a constant density
of states $\Gamma_P(\omega) \sim \Gamma, \; \forall \omega $.
 The Fermi function 
$f(\omega)= 1/(e^{\beta (\omega - \mu)} + 1)$,
depends on the chemical potential $\mu$ of the $L$ and $R$ reservoirs, which
we take as a reference and on the temperature $1/\beta$ that we assume
to be same  for all the reservoirs.  
The retarded and lesser Green functions
are, respectively, evaluated by solving Dyson equations:
\begin{eqnarray}\label{dy}
& & -i \partial_{t^{\prime}} \hat{G}^R(t,t^{\prime})- \hat{G}^R(t,t^{\prime})\hat{H}_{sys}(t) \nonumber \\
& & -
\int dt_1\hat{G}^R(t,t_1) \hat{\Sigma}^R(t_1,t^{\prime}) =\hat{1} \delta(t-t^{\prime}), \\ 
& & \!\!\! \hat{G}^<(t,t^{\prime})= \int dt_1 dt_2 \hat{G}^R(t,t_1) \hat{\Sigma}^<(t_1,t_2)
\hat{G}^A(t_1,t^{\prime}),\nonumber
\end{eqnarray}
with $\hat{G}^A(t,t^{\prime})= [\hat{G}^R(t^{\prime},t)]^{\dagger} $ and $\hat{\Sigma}^A(t,t^{\prime})= 
[\hat{\Sigma}^R(t^{\prime},t)]^{\dagger} $.
The elements of the Green function matrices are defined over the sites of the
central device and $\hat{1}$ denotes the identity matrix in this space. Similarly, the Hamiltonian matrix
$\hat{H}_{sys}(t)$ contains the matrix elements of the Hamiltonian
(\ref{hsys}), while
 the self-energy matrix contains non-vanishing elements only
at the sites $l_{\alpha}$, that are in contact to the reservoirs $\alpha=L,R,P$.

As in previous works \cite{lilip,lilimos,us}, we introduce the following representation
for the retarded Green function:
\begin{eqnarray}\label{hatg}
\hat{G}^R(t,t^{\prime})&=& \sum_k \int \frac{d \omega}{2 \pi} e^{-i k \Omega_0 t}
e^{-i \omega (t-t^{\prime})} \hat{\cal G}(k,\omega).
\end{eqnarray}
In addition, for weak driving voltages, it is natural to assume
also weak amplitudes for the dc and ac voltages of the probe. Thus, the exponential of Eq. (\ref{phi})
 simplifies to:
\begin{eqnarray}\label{phi}
\phi(t,t^{\prime}) & \sim  & 
\{ 1 - i [ (\mu_P - \mu)(t-t^{\prime}) \nonumber \\
& & + \int^t_{t^{\prime}} dt_1 V_P \cos( \Omega_0 t_1 + \varphi_P )] \}.
\end{eqnarray}
Introducing  expressions Eqs.(\ref{hatg}) and (\ref{phi}) in Eq.(\ref{jpt}), taking into account 
the assumption of a low driving
frequency by keeping terms up to ${\cal O}(\Omega_0)$, and keeping terms up to
${\cal O}( w_P)$ due to the non-invasiveness of the probe,  we arrive, after some algebra,
 to the following expression for current flowing through the contact between
the central system and the probe:
\begin{eqnarray}\label{jptf}
J_P(t) & = & \sum_{k=-1}^1
\sum_{\alpha=L,R,k^{\prime}} e^{-i k\Omega_0 t }
\int \frac{d \omega}{2 \pi} \frac{\partial f(\omega)}{\partial \omega} \Gamma_{\alpha}(\omega)  
\Gamma_{P}(\omega)
\nonumber \\
& & \times
{\cal G}_{l_{P},l_{\alpha}}(k+ k^{\prime},\omega) [{\cal G}_{l_{P},l_{\alpha}}(k^{\prime},\omega) ]^*
\times 
\nonumber \\
& &[V_P \cos(\Omega_0 t + \varphi_P) -\mu + \mu_P - k^{\prime} \Omega_0] - i \sum_ {k=-1}^1 k \Omega_0 
\nonumber \\
& &\times e^{i k \Omega_0 t} \int \frac{d \omega}{2 \pi}  
\frac{\partial f(\omega)}{\partial \omega}
[{\cal G}_{l_{P},l_P}(k,\omega) ]^* \Gamma_{P}(\omega).
\end{eqnarray}
As discussed in Refs. \onlinecite{mobup,lilimos}, it is possible to split the time-dependent 
current into  different
components:
\begin{equation}
J_P(t)=J^{pump}(t)+ J^{bias}(t).
\end{equation}
The first one corresponds to pure pumping processes and behaves like $J^{pump}(t) \propto \Omega_0$, 
while the other one is the contribution due to the existence of a bias, and behaves like
 $J^{bias}(t) \propto U_P(t)$. In this approximation we are
neglecting the interference term, which is $\propto \Omega_0 U_P(t)$. In terms of the driving parameters,
the latter term contributes at ${\cal O}(\Omega_0 U_0^4)$ to the dc component and
at ${\cal O}(\Omega_0 U_0^3)$ to the first harmonic of the probe voltage.

For weak driving, the Dyson equation Eq. \ref{dy} can be solved perturbatively. \cite{lilip} 
The terms necessary to evaluate the conditions of Eq. (\ref{cond}) exactly 
up to ${\cal O}(U_0^2)$, ${\cal O}(\Omega_0)$
and ${\cal O}(w_P)$ are:
\begin{equation} \label{pert}
\hat{G}(t,\omega) \sim  \sum_{k=-1}^1  \hat{\cal G}(k,\omega) e^{-i k \Omega_0 t},
\end{equation}
with
\begin{eqnarray} \label{pert1}
\hat{\cal G}(0,\omega) & \sim &  \hat{G}^0(\omega), \nonumber \\
 \hat{\cal G}(\pm 1,\omega)  & \sim & \hat{G}^0(\omega) \hat{V}^{(\pm 1)} \hat{G}^0(\omega),
\end{eqnarray}
where $\hat{G}^0(\omega)$ is the equilibrium retarded Green function of the central system
described by the Hamiltonian $H_0$, coupled only to the $L$ and $R$ reservoirs, while:
\begin{equation}
V^{(\pm 1)}_{l,l^{\prime}} = \frac{U^0_{l,l^{\prime}}}{2} e^{\mp i \varphi_{l,l^{\prime}}}.
\end{equation}
Inserting these functions into in the time-dependent current Eq. (\ref{jptf}) and imposing the conditions 
Eq. (\ref{cond}), we obtain the following set of linear coupled equations
 that must be fulfilled  in order to have  a vanishing time-dependent current:
\begin{eqnarray} 
\mu-\mu_P &=& \Omega_0 \lambda^{(0)}_P + \mbox{Re}\{ V_P e^{i \varphi_P} \lambda^{(1)}_P\}, 
\label{deltamu} \\
 \frac{V_P}{2} e^{-i \varphi_P} &=&  \Omega_0 \lambda^{(2)}_P+ (\mu-\mu_P) \lambda^{(1)}_P, \label{vp}
\end{eqnarray}
being
\begin{eqnarray}
\lambda^{(0)}_P & = & - \frac{ \sum_{\alpha} \sum_{k=\pm 1} k |{\cal G}_{l_P,l_{\alpha}}(k,\mu)|^2 \Gamma_{\alpha}(\mu)
}{ \Lambda_P }, \nonumber \\
\lambda^{(1)}_P & = & \frac{ \sum_{\alpha} \sum_{k=-1}^{0} {\cal G}_{l_P,l_{\alpha}}(k+1,\mu) [{\cal G}_{l_P,l_{\alpha}}(k,\mu)]^* 
 \Gamma_{\alpha}(\mu)}
{ \Lambda_P} ,\nonumber \\
\lambda^{(2)}_P & = & - \frac{ \sum_{\alpha} {\cal G}_{l_P,l_{\alpha}}(0,\mu)[{\cal G}_{l_P,l_{\alpha}}(-1,\mu)]^*\Gamma_{\alpha}(\mu)}
{\Lambda_P}\nonumber \\
& & 
-i \frac{ [{\cal G}_{l_P,l_P}(-1,\mu)]^* }{\Lambda_P },
\nonumber \\
\Lambda_P &=& \sum_{\alpha} |G^0_{l_P,l_{\alpha}}(\mu)|^2 \Gamma_{\alpha}(\mu),
\end{eqnarray} 
where the sum in $\alpha$ runs over $L,R$ and, for simplicity, we have assumed zero temperature.
 It is interesting to notice in the above equations, that the very
existence of an ac component in the voltage probe ($V_P$) modifies the dc component of the voltage
profile $\mu_P-\mu$. 

Finally, keeping only terms up to ${\cal O }(U_0^2)$ in the solution of 
Eqs. (\ref{deltamu},\ref{vp}), we obtain the dc and ac components of the voltage profile 
sensed by the voltage probe. They respectively
read:
\begin{eqnarray} 
\mu-\mu_P & = &\Omega_0
\{ \lambda^{(0)}_P + \frac{1}{2}\mbox{Re} [\lambda^{(1)}_P  (\lambda^{(2)}_P)^*] \}, \label{deltamuf}  \\
 V_Pe^{-i \varphi_P} &=&  2 \Omega_0 \lambda^{(2)}_P, \label{vpf}
\end{eqnarray}
where dc the component is  $\propto \Omega_0 U_0^2$, while the ac one is  $\propto \Omega_0 U_0 $.

At this point it is important to compare Eq. (\ref{deltamuf}) with the  result obtained
for the case of a dc probe, as  the one  considered in Ref. \onlinecite{us}. 
The latter case corresponds to take $V_P=0$,
in the above expressions, which in turn leads to a dc voltage profile $\mu_P-\mu$ given only by the
first term of Eq. (\ref{deltamuf}). It is easy to verify that this result 
coincides with Eq. (16) of our previous work. \cite{us,note}
The dc voltage probe senses scattering events 
at static barriers, like walls and impurities of the structure, as well as at the dynamical 
pumping centers, with the characteristic that they take place within the same
Floquet channel. These  processes are contained in the first term ($\lambda^{(0)}_P$) of 
the above expressions. On the other hand, the ac component of the voltage profile 
senses additional effects due to scattering processes between different
Floquet components. These are collected into  the terms $\lambda^{(1)}_P$ and $\lambda^{(2)}_P$
of Eq.(\ref{deltamuf}).
 The remarkable  and new feature that the
ac-dc voltage probe brings about is  contained in  these inter-Floquet scattering processes
that lead to   a correction of the same order in the driven parameters, i.e. an extra 
term $\propto \Omega_0 U_0^2$, that has to be added to the result obtained with a dc voltage probe.

\subsection{Four terminal impedance} \label{z4}
The  dc
component of the current entering the  $L$ and $R$ reservoirs
satisfies the relation $J_L^{(0)}=-J_R^{(0)}$. However, the
higher harmonics $J_{\alpha}^{(\pm 1)}$ are not expected to satisfy
such a condition (see Ref. \onlinecite{tdep2_noise1}). This is because there may be
instantaneous accumulation of charge with vanishing average
along the structure.
 Therefore, 
in order to define the impedance of the device, we choose the
current flowing through the left contact 
 as a reference. 

For weak driving, the time-dependent currents (\ref{tdepcur}) flowing into the $L$
and $R$ contacts have the following components:
\begin{eqnarray}
J_{\alpha}^{(0)} & = &\Omega_0 
\sum_{\beta=L,R} \sum_{k}  \Gamma_{\alpha}(\mu)
 \Gamma_{\beta} (\mu) \nonumber \\
& & \times k 
|{\cal G}_{l_{\alpha}, l_{\beta}}(k, \mu)|^2, \nonumber \\
J_{\alpha}^{(\pm 1)} & = &  
 \Omega_0  
\sum_{\beta=L,R} \sum_{k=\mp 1,0}  \Gamma_{\alpha}(\mu)
 \Gamma_{\beta} (\mu) \nonumber \\
& & \times k
{\cal G}_{l_{\alpha}, l_{\beta}}(k  \pm 1, \mu)  
{\cal G}_{l_{\alpha}, l_{\beta}}(k, \mu)^* 
\nonumber \\
& &  \mp i   
{\cal G}_{l_{\alpha}, l_{\alpha}}(\mp 1, \mu)^* \Gamma_{\alpha}(\mu) \},
\end{eqnarray}
being $J_{\alpha}^{(1)}=[J_{\alpha}^{(-1)}]^*$,
with the Green functions given by (\ref{pert}) and (\ref{pert1}).

In a four terminal measurement, with ac-dc non-invasive probes,  the
voltage drop between the points $l_P$ and $l_{P^{\prime}}$ is simply the difference
of the voltage sensed by each probe, regarding each of them as independent of one another, thus:
\begin{equation}
\Delta V (t) = \sum_{k=-1}^1 \Delta V^{(k)} e^{-i k \Omega_0 t},
\end{equation}
being the ac and dc components, respectively:
\begin{eqnarray}
\Delta V^{(\pm 1)} & = & 
V_P e^{\pm i \varphi_P}- V_{P^{\prime}} e^{\pm i \varphi_{P^{\prime}}},\nonumber \\
\Delta V^{(0)} & = & \mu_P - \mu_{P^{\prime}}.
\end{eqnarray}

Thus, the  four-point impedance also has ac and dc components defined as follows:
\begin{equation}
Z^{(k)}=\frac{ \Delta V^{(k)} }{J_L^{(k)}}, k= -1,0,1,
\end{equation}
being $Z^{(1)}=[Z^{(-1)}]^*$.
In linear circuits with ac and dc sources, the dc component of the impedance,
$Z^{(0)}$ is simply the resistance,
and it coincides with the real part of $ Z^{(1)}$. In our quantum case, we cannot provide
any proof on the validity of
such a relation between the two components of the impedance, and we must simply 
regard them as providing 
different pieces of information about the driven system.

Of particular interest is the behavior of $Z^{(0)}$, which should be regarded as an
extended definition of the dc four-point resistance $R_{4t}$ we have introduced in
Ref. \onlinecite{us}. In the next section we present results for a particular
model of driven system. In general, we can notice that the four-terminal resistance
sensed by dc probes is associated with the first term of Eq. (\ref{deltamuf}):
\begin{equation}
R_{4t}= \Omega_0
\frac{ \lambda^{(0)}_{P^{\prime}} -  \lambda^{(0)}_{P} }{J_L^{(0)} }.
\end{equation}
The dc impedance contains an additional term, which is of the
same order of magnitude,  associated to scattering
processes mediated by the ac voltage of the probe:
\begin{equation}
Z^{(0)}=R_{4t}+  \Omega_0
\frac{ \mbox{Re} [\lambda^{(1)}_{P^{\prime}}  (\lambda^{(2)}_{P^{\prime}})^*
-\lambda^{(1)}_{P}  (\lambda^{(2)}_{P})^*]  }{2 J_L^{(0)} }.
\end{equation}

\section{Results for a simple model of a quantum pump}\label{s2}
We now examine the concepts introduced in the previous section in the context
of a quantum pump device. We consider a simple model where two ac gate potentials with the
same amplitude $V_0$ oscillate with a phase-lag $\delta$ at two barriers confining 
a quantum dot. The dot and barriers are described by the Hamiltonian  $H_0$ introduced in
section \ref{model}, with hopping
between nearest neighbor positions on a one-dimensional lattice of $N$ sites, and barriers at the
positions $l_A$ and $l_B$ of that lattice:
\begin{equation} \label{eps}
\varepsilon_{l,l'}= \delta_{l,l'}[\delta_{l,l_A}+\delta_{l,l_B}] E_B.
\end{equation}
The driving terms read:
\begin{equation}
U_{l,l'}= V_0 \delta_{l,l'}[\cos(\Omega_0 t + \delta)\delta_{l,l_A} +
 \cos(\Omega_0 t )\delta_{l,l_B}].
\end{equation}

\begin{figure}
\includegraphics[width=0.9\columnwidth,clip]{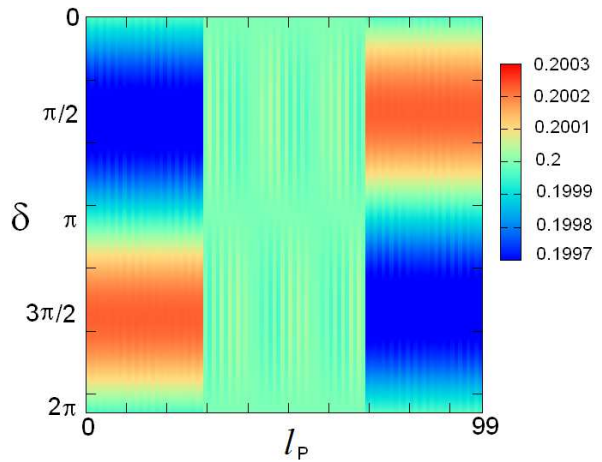}
\caption{\label{fig2} (Color on-line) Contour plot for the 
local dc component of the voltage $\mu_P$  sensed by the ac-dc voltage probe  as  
function of the probe position $l_P$   along the system (horizontal axis) and the
phase lag $ \delta$. We consider a
a quantum pump modeled by a driven chain with
by $N=99$ sites with two  barriers of height $E_B=0.2$ located at  $l_A=30$ and $l_B=70$
 as indicated by the vertical dashed lines.
 The chemical potential is $\mu=0.2$, which corresponds to $k_F =1.47$. }
\end{figure}

\begin{figure}
\includegraphics[width=0.9\columnwidth,clip]{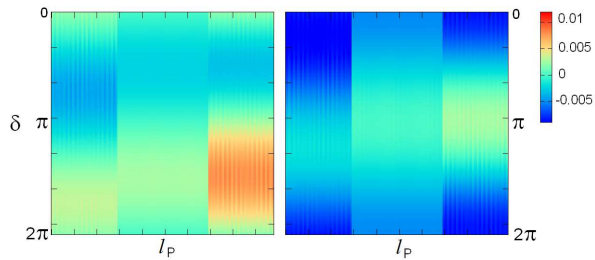}
\caption{\label{fig3} (Color on-line) Contour plot for the real (left panel)
and imaginary part (right panel) of the ac component of the local potential
$V_P e^{i \varphi_P}$   sensed by the ac-dc voltage probe  as  
function of the probe position $l_P$   along the system (horizontal axis) and the
phase lag $ \delta$. Other parameters are the same as in Fig. \ref{fig2}.}
\end{figure}

The behavior of the dc and ac components of the voltage profile are shown, respectively,
in Figs. \ref{fig2} and \ref{fig3}, as functions of the position of the structure at
which the probe is connected, $l_P$,  and the phase-lag 
$\delta$ of the pumping potentials. 
In the case case of the ac component, 
we plot separately the behavior of the real and imaginary part of $V_P e^{i \varphi_P}$.

As discussed in section \ref{volt}, the ac-dc voltage probe senses a profile
which significantly differs from the one sensed by a pure dc probe like the
one we have considered in Ref. \onlinecite{us}. This is because, the dc probe
measures just the scattering processes that take place within a single Floquet
channel, which are described by only the first term of Eq. (\ref{deltamuf}).
Instead, the additional ac components of voltage of the probe, 
$V_P^{\pm i \varphi_P}$, mediate scattering processes between different
Floquet channels. The consequence is that the dc component of the
profile sensed by the ac-dc probe contains, in addition, the second term
of Eq. (\ref{deltamuf}), which is of the same order of magnitude as the first one.

\begin{figure}
\includegraphics[width=0.9\columnwidth,clip]{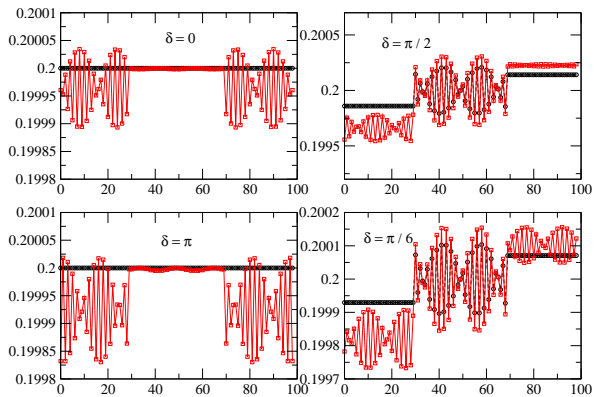}
\caption{\label{fig4} (Color on-line) The 
local dc component of the voltage $\mu_P$  sensed by the ac-dc voltage probe  as  
function of the probe position $l_P$   along the system for different phase lags
$ \delta$, which are indicated in the Fig. 
In dashed lines, we indicate the profile corresponding to sensing with a dc
probe.
Other parameters are the same as in Fig. \ref{fig2}. }
\end{figure}

In order to make clear the difference between the two procedures of defining
the dc component of the voltage profile, we plot the prediction of  Eq. (\ref{deltamuf})
for a set of representative values of the 
phase-lag $\delta$ in Fig. \ref{fig4}, also showing  in dashed lines the prediction
obtained from a pure dc probe like the one of Ref. \onlinecite{us}. We notice that
several interesting features can be identified in the  voltage landscape
 for the case of the general ac-dc probe. 

A first point worth of mention is that for
the present system, which contains two pumping potentials with the same amplitude, the dc voltage
profile sensed by a dc probe is flat along the structure and equal to zero
when $\delta= 0, \pi$. This behavior goes in line with the behavior of
the dc-current along the structure, which vanishes for these values of the phase-lag
as a consequence of the symmetries of the system. \cite{lilisim} Moreover,
 it can be shown that for weak driving the dc current in this model
behaves like $J^{0} \propto \Omega_0 (V_0)^2 \sin(\delta)$, \cite{brou0,lilip}
and that for a fixed position of the probe, the dc voltage sensed
by the dc probe follows exactly this behavior as a function of $\delta$. \cite{us}  
In the case of the probe containing the additional ac voltage, the dc and ac components
of the current along the structure are not affected by the probe, provided that it is 
weakly coupled. However, the additional inter-Floquet scattering processes mediated 
by the probe
contribute to break symmetries and the dc profile is no longer an odd
function of $\delta$, displaying non-vanishing features
for $\delta=0, \pi$, as shown in Figs. \ref{fig2} and \ref{fig4}. Our results
show that the dc profile remains invariant under the following simultaneous 
transformations: 
$\delta \rightarrow -\delta; \;\; x \rightarrow -x $, where the latter
operation denotes a spatial inversion with respect to the center of the structure
(see Fig.  \ref{fig2}). On the other hand,  the analysis
of the ac component of the voltage shown in Fig. \ref{fig3} cast  the invariance
of the amplitude $V_P$ under   
the simultaneous transformation:
$\delta \rightarrow -\delta; \;\; x \rightarrow -x; 
\;\; \varphi_P \rightarrow \varphi_P + \delta $. These symmetry properties are rather
 expected and fully consistent with 
 the symmetry properties of the structure we are studying. However,
we stress the remarkable
fact that the dc voltage profile does not follow as a function of $\delta$
 the behavior of the dc component
of the current, as it is the case of the one defined from the pure dc probe. 
In the simple one-channel model we are considering, we cannot analyze the
symmetry properties of the voltage profile in the presence of a magnetic
field. In general, in the presence of a magnetic flux $B$, the non-interacting
Green function satisfies the following symmetry: $\hat{G}^0(B, \omega) \rightarrow
[\hat{G}^0(-B, \omega)]^t$, where the superscript $t$ denotes the transposed matrix. 
The way in which this transformation affects the voltages (\ref{deltamu}) and
(\ref{vp}) is not obvious and is expected to be model-dependent. This is also
the case of the dc pumped current, as discussed in previous works. \cite{mobum, brou0} 

\begin{figure}
\includegraphics[width=0.9\columnwidth,clip]{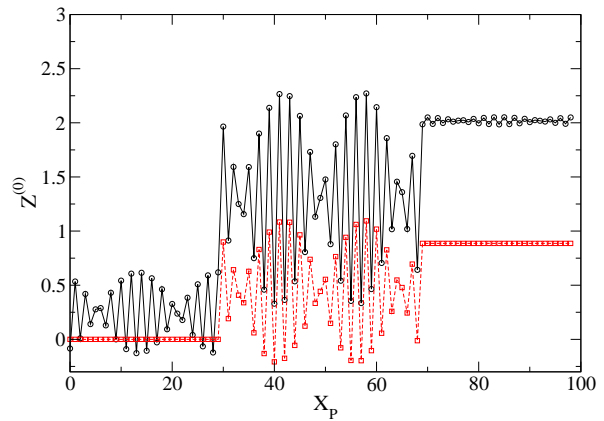}
\caption{\label{fig4p} (Color on-line) The real (left) and imaginary (right) 
local ac component of the voltage $V_P e^{i \varphi_P}$  sensed by the ac-dc voltage probe  as  
function of the probe position $l_P$   along the system for different phase-lags
$ \delta=0, \pi/2, \pi$, (top to bottom).
Other parameters are the same as in Fig. \ref{fig2}. }
\end{figure}

\begin{figure}
\includegraphics[width=0.9\columnwidth,clip]{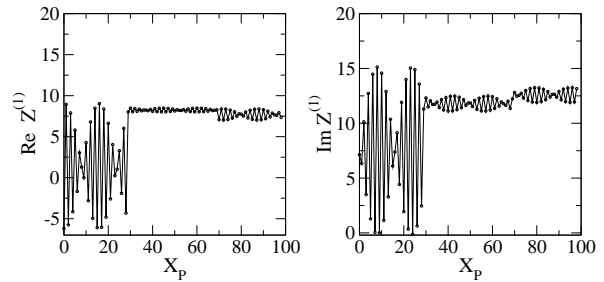}
\caption{\label{fig5} (Color on-line) The 
dc component of the impedance $Z^{(0)}$, connecting the first probe at
$l_P=10$,
 as a function of the position of the second probe
$l_{P^{\prime}}$  for 
$ \delta = \pi/2 $. The four terminal resistance $R_{4t}$ determined 
by a dc probe is also plotted in dashed lines.
 Other parameters are the same as in Fig. \ref{fig2}.}
\end{figure}

Another feature worth of notice is the pattern distinguished as a sequence
of fringes in Figs. \ref{fig2} and \ref{fig3} and as oscillations of the dc and ac 
voltage landscapes of Figs. \ref{fig4} and \ref{fig4p}.
 The ultimate origin of these features is the existence of Friedel oscillations.
For this reason, they show the characteristic spatial period of $2 k_F$. In fact, it can be verified that 
the function $A \sin(2 k_F l_P+\alpha)$, with suitable factors $A$ and $\alpha$, and 
$l_P$ adopting integer values (recall that we are studying a lattice model), displays
the same plot as the oscillatory part of the plots shown in this figure, for the considered
value of $k_F$.
 The
sources for these oscillations are the different scattering centers of static and dynamical character.
In particular, each of the two barriers of the structure, defined in the static profile
(\ref{eps}), behaves like a static impurity which generates
 usual Friedel oscillations.\cite{gram}
On top of this, we also have dynamical scattering centers due to the pumping potentials. 
The consequence is the generation of $2 k_F$ oscillations
in both dc and ac components of the local density of states along the system.\cite{us}
The ensuing scattering processes being encoded within the different
components of the Green function $G(k,\omega), k=-1,0,1$ and are detected in
the dc as well as in the ac components of the voltage probe.
The oscillations generated at the 
different sources, interfere and may become vanishingly small within some regions of
the sample, depending on the value of the phase-lag $\delta$ (see, for example the
central region between the two barriers in the two left-hand panels of Fig. \ref{fig4}
and in the top panels of Fig. \ref{fig5}). Notice that, besides the oscillations,
the ac-voltage at the probe leads to a dc voltage drop that can be as
large as twice the magnitude defined by a pure dc probe. This feature is important
since in real experiments like that of Ref. \onlinecite{SMCG99}, it is the voltage
drop the measured quantity from where the pumped dc current is inferred. The
dc voltage drop sensed by a pure dc probe is related to this current by a simple
multiplicative factor, the dc four point resistance $R_{4t}$, which is expected to
depend only on the properties of the sample. However, the  presence 
of an ac voltage at probe can increase the dc voltage drop in a way that when simply
dividing this quantity by $R_{4t}$, the inferred dc current  can significantly
differ from the actual one. In Refs. \onlinecite{brou,mobup} it has been pointed out that
the presence of voltage probes could increase the intensity of the dc current through the 
structure, by adding to the pumped current a contribution due to rectification mechanisms.
Our results suggest that the dc current through the structure may remain 
unaffected by the probe, but when its magnitude is obtained by naively dividing 
the dc voltage drop by some estimate for $R_{4t}$, we could conclude that it is much 
larger than its actual value.

\begin{figure}
\includegraphics[width=0.9\columnwidth,clip]{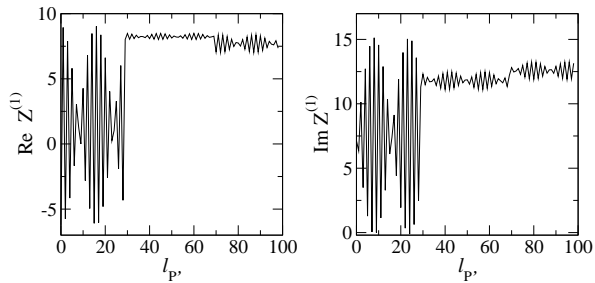}
\caption{\label{fig6} (Color on-line) Real (left panel) and imaginary (right panel) parts of the 
ac component of the impedance $Z^{(1)}$, connecting the first probe at
$l_P=10$,
 as function of the position of the second probe
$l_{P^{\prime}}$  for 
$ \delta = \pi/2 $, which are indicated in the Fig. 
 Other parameters are the same as in Fig. \ref{fig2}.}
\end{figure}

Finally, we show in Figs. \ref{fig5} and \ref{fig6}, respectively, the dc  and
ac components of the impedance for a particular value of the phase-lag. As discussed 
in section \ref{z4}, the 
dc component $Z^{(0)}$ differs from the dc four-terminal resistance $R_{4t}$ by the
term due to scattering processes mediated by the ac potential of the probe. In Ref.
\onlinecite{us} we have shown  for the present simple model of a quantum
pump, that $R_{4t}$ of the full device, defined from a dc four terminal measurement with
probes connected outside the driven region, 
is a universal property of the system, independent of the driving mechanism.
 Namely,  it coincides with the one
defined when the transport is induced by an equivalent dc voltage applied between the
$L$ and $R$ reservoirs.  On the contrary, the impedances $Z^{(0)}$ and $Z^{(1)}$, strongly
depend on the positions where the probes are connected and account for the spacial 
oscillations
of the voltage profile. Irrespectively of the position of the probes, for a given
$\delta$, the behavior
of the impedances along the structure contain  the $2k_F$ oscillations of the voltage
profiles.

\section{Summary and conclusions}\label{con}
In this work we have introduced the theoretical concept of the ac-dc voltage probe, as a 
weakly coupled
reservoir with a  time-dependent voltage which instantaneously 
adapt to fulfill the local equilibrium condition of both  dc and time-dependent current 
flowing between the driven system and the probe. 
We have focused on the weak driving regime, where the currents, as well
as the probe voltage contain a single harmonic on top of their dc components. The procedure
we have introduced, can be generalized to consider stronger driving and include
additional harmonics. Under the assumption of  non-invasive probes, the 
information of the voltage drop is enough to define the four-point impedance.

We have found that the dc component of the voltage defined in this way, differs 
 from the one defined by a pure dc voltage probe. In particular,
the ac-dc probe is able to capture scattering processes between different Floquet
channels that are not detected by the pure dc one. These additional processes are of the
same order of magnitude as the inter-Floquet ones and may
introduce relevant qualitative features in the behavior of the voltage profile.
In the particular case of a quantum pump with two ac potentials oscillating with a
 phase-lag,
the dc voltage profile does not follow the same functional behavior with the 
phase-lag observed in the dc current. This feature also plagues  the behavior
of the dc component of the impedance $Z^0$. As a consequence, unlike the dc four point
resistance $R_{4t}$, the dc impedance 
is not a universal quantity which depends on just the geometrical properties of
the structure, but also depends on the position at which the voltage probes are 
connected.

One of the important messages of these theoretical ideas towards the experimental realm 
is related to the inference of the behavior of the dc current induced in the quantum
pump from a four-terminal voltage measurement. The relation between these two quantities 
is not a simple factor as it could be naively expected. In Refs. 
\onlinecite{brou} and \onlinecite{mobup} it was discussed that the ac potentials
of the voltage probes used in four terminal measurements in quantum pumps, as 
in the experiment
by Switkes and coworkers \cite{SMCG99}, could act as additional sources and result in a dc
 current higher than the one induced by bare pumping potentials. In this work, we 
have considered  non-invasive probes, which do not induce additional
rectified currents through the structure. We
have, however, shown that they anyway introduce additional scattering processes with the outcome
of additional features in  the  sensed voltage landscape.

\section{Acknowledgements}
We thank M. J. Sanchez for many discussions. We acknowledge support from CONICET,
PIP 1212-09, UBACYT, Argentina, and J. S. Guggenheim Foundation (LA).


\begin{thebibliography}{99}
\bibitem{note}There is a typo in Eq. (16) of Ref. \onlinecite{us}. 
The factor $(V/2)^2$ should be omitted.
\end{thebibliography}
\end{document}